\documentclass[aps,pre,floats,twocolumn,superscriptaddress,longbibliography]{revtex4-1}

\usepackage{natbib}
\usepackage[utf8]{inputenc}

\usepackage{amsmath, amsthm, amssymb}
\usepackage{enumitem}
\usepackage{graphics,graphicx}% Include figure files
\usepackage{epsfig}
\usepackage{times}% Times fonts
\usepackage{dcolumn}% Align table columns on decimal point
\usepackage{url}
\usepackage{color}
\usepackage{wasysym}

\usepackage[all]{xy}
\usepackage{mdwlist}

\definecolor{snorkelBlue}{rgb}{0,0.31,0.52}
\definecolor{peachColor}{rgb}{0.97,0.47,0.42}
\definecolor{Biscaybay}{rgb}{0, 0.556, 0.655}
\definecolor{lightGray}{rgb}{0.7,0.7,0.7}

\usepackage[all]{xy}
\usepackage{mdwlist}
\usepackage[normalem]{ulem}

\newcommand{\beq}{\begin{equation}}
\newcommand{\eeq}{\end{equation}}
\newcommand{\bea}{\begin{eqnarray}}
\newcommand{\eea}{\end{eqnarray}}

\newcommand{\Snp}{S_{\mbox{\scriptsize np}}}
\newcommand{\Inp}{I_{\mbox{\scriptsize np}}}
\newcommand{\Sp}{S_{\mbox{\scriptsize p}}}
\newcommand{\Ip}{I_{\mbox{\scriptsize p}}}
\newcommand{\Tpnp}{\Gamma_{\mbox{\scriptsize p$\to$np}}}
\newcommand{\Tnpp}{\Gamma_{\mbox{\scriptsize np$\to$p}}}
\newcommand{\qnp}{q_{\mbox{\scriptsize np}}}
\newcommand{\qp}{q_{\mbox{\scriptsize p}}}
\newcommand{\Pp}{P_{\mbox{\scriptsize p}}}
\newcommand{\Pnp}{P_{\mbox{\scriptsize np}}}
\newcommand{\ie}{\emph{i.e.} }

\begin{document}

%%--------------------------------%%
%%           TITLE                  %%
%%--------------------------------%%
\title{Pulsating-campaigns of human prophylaxis driven by risk perception palliate oscillations of direct contact transmitted diseases}

%%------------------------------------------------------------%%
%%            AUTHORS & AFFILIATIONS              %%
%%------------------------------------------------------------%%
\author{Benjamin Steinegger}
\affiliation{Departament d'Enginyeria Inform\`atica i Matem\`atiques, Universitat Rovira i Virgili, E-43007 Tarragona, Spain}

\author{Alex Arenas}
\affiliation{Departament d'Enginyeria Inform\`atica i Matem\`atiques, Universitat Rovira i Virgili, E-43007 Tarragona, Spain}

\author{Jes\'us G\'omez-Garde\~nes}
\affiliation{Department of Condensed Matter Physics, University of Zaragoza, E-50009 Zaragoza, Spain}
\affiliation{GOTHAM Lab -- Institute for Biocomputation and Physics of Complex Systems (BIFI), University of Zaragoza, E-50018 Zaragoza, Spain}

\author{Clara Granell}
\affiliation{Department of Condensed Matter Physics, University of Zaragoza, E-50009 Zaragoza, Spain}
\affiliation{GOTHAM Lab -- Institute for Biocomputation and Physics of Complex Systems (BIFI), University of Zaragoza, E-50018 Zaragoza, Spain}

%%---------------------------------------%%
%%            ABSTRACT                %%   % 145 words
%%---------------------------------------%%
%A normal abstract, 500 words max AND 5% of total length, unreferenced.
\begin{abstract}
Human behavioral responses play an important role in the impact of disease outbreaks and yet they are often overlooked in epidemiological models. Understanding to what extent behavioral changes determine the outcome of spreading epidemics is essential to design effective intervention policies. Here we explore, analytically, the interplay between the personal decision to protect oneself from infection and the spreading of an epidemic. We do so by coupling a decision game based on the perceived risk of infection with a Susceptible-Infected-Susceptible model. Interestingly, we find that the simple decision on whether to protect oneself is enough to modify the course of the epidemics, by generating sustained steady oscillations in the prevalence. We deem these oscillations detrimental, and propose two intervention policies aimed at modifying behavioral patterns to help alleviate them. Surprisingly, we find that pulsating campaigns, compared to continuous ones, are more effective in diminishing such oscillations.
\end{abstract}

%\pacs{%
%89.65.-s,	%Social and economic systems
%89.75.Fb,	%Structures and organization in complex systems
%89.75.Hc  %Networks and genealogical trees
%}

\maketitle

%%---------------------------------------%%
%%                INTRO                   %%   Currently: approx < 728 words
%%---------------------------------------%%
\section{Introduction}

 In a context of a disease outbreak, it is known that humans might change their behavioral patterns in an effort to avoid infection~\cite{Hays2006}. Behavioral responses range from wearing face masks or increasing hand hygiene to prevent influenza-like illnesses, to using condoms to stop sexually transmitted diseases, to avoid traveling to an infection locus, among others. In the past years, substantial efforts have been made by governments and policy makers to monitor the spreading of diseases and implement effective containment measures, some of which include recommending certain behavioral changes to the population. Though crucial, the effect of changing human behavior is often overlooked in epidemiological models for disease spreading that might ultimately be used for policy-making. Studying the effect of individual human responses to the presence of infectious diseases in a population is of outmost importance as it is known to alter the spreading dynamics and can cause a systematic bias in the disease forecast if not taken into account~\cite{Eksin2018,Payn1997}.

Reports of the existence of an effect of human behavior on the spreading of epidemics have been known for long. However, it has not been until recently that some mathematical models have explicitly incorporated the effect of individual behavioral responses on the outcome of an epidemic~\cite{Funk2010Review,Wang2016}. Some interesting mathematical models consider, for example, individuals lowering their daily contact activity rates once an epidemic has been identified in a community~\cite{delValle2005,DOnofrio2009,Moinet2018}, individuals rewiring their contacts with infected neighbors~\cite{Gross2006,Althouse2014,Scarpino2016,Sherborne2018},
 or the selfish behavior of individuals confronting vaccination campaigns and its effect on the endemicity of the epidemics~\cite{Bauch2004,reluga2006,donofrio2007,vardavas2007,Fu2011,Cardillo2013,Steinegger2018}.

A very common way of factoring in the effect of human behavior in epidemics is to consider that individuals alter their behavior according to the prevalence level of the disease. Indeed, empirical studies~\cite{ahituv1996, philipson1996} have shown that protective behavior increases as a disease becomes more prevalent. However, some surveys~\cite{anderson1999condom, prata2006relationship} indicate that individuals base the decision on whether to protect themselves in their {\em perceived} risk of infection, which may differ from their real risk of infection as other factors are taken into account. These factors may include one's perceived susceptibility, the number of reported cases of diseased individuals, the distance to the focus of the epidemic, the cost of the measures to be taken, etc. 

In our model, we assume that individuals act according to their perception of risk of infection, and more particularly, that individuals will be inclined to take preventive measures if their perceived risk exceeds the cost associated to taking the prophylactic measure. 
We formulate the decision problem as a two-strategy game theory dilemma, where the individual's perceived risk depends on the cost of contracting the disease and its prevalence. One strategy is to take protective measures against the disease, in which case we say that the agent is {\em protected} (P); the other strategy is to disregard any behavioral change and stay {\em not protected} (NP). Taking the protective measures will imply, in terms of the epidemics, that this individual will have a lower chance of getting infected if she is susceptible, and that she will be less infectious to others in case she is infected. We couple this decision game with an epidemic process, modeled using the Microscopic Markov Chain Approach~\cite{MMCA}, for the SIS model ---in which agents can either be {\em susceptible} (S) or {\em infected} (I)--- and we let both  processes evolve simultaneously. On the one hand, the decision on whether to adopt protective measures is made according to the risk perceived in each strategy according to global information. On the other hand, the epidemics propagates in a contact network, using then local information. We call this model ``risk--driven epidemic spreading'' given that the epidemics is palliated by the individual prophylaxis which in turn is driven by their risk perception.

We will present our analysis for surrogates of contact direct networks.
%(see Methods: Network model). 
First, we investigate mathematically the nonlinear interplay between the risk-perception decision and the prevalence of the disease, whose outcome is a sustained oscillation in time. We scrutinize the role of the different parameters of the model with particular focus on the effectiveness of the protection method, that plays a key role in the oscillations. Furthermore, we provide the exact epidemic thresholds and protection thresholds. Finally, we evaluate two types of awareness campaigns leveraging the full predictive power of the model: a continuous awareness campaign that is active through time, and a pulsating campaign that is activated only when the epidemics is on the rise. The results prove that pulsating campaigns are more effective to contain the prevalence of the disease.

%%---------------------------------------%%
%%              MODEL                    %%
%%---------------------------------------%%  < 784w (no compto eqs)
\section{Risk perception-driven epidemic spreading model} \label{sec:model}

In our risk--driven epidemic spreading model, agents can be in four possible states: Infected and protected, infected and not protected, susceptible and protected, or susceptible and not protected. Formally speaking, let us define the macroscopic quantities $\Ip(t)$, $\Inp(t)$, $\Sp(t)$, and $\Snp(t)$ as the fraction of the population in each of the former possible states at time $t$, respectively. 

The agents will choose one of the two possible strategies according to the difference of payoffs of each strategy. The payoff of being protected ($P_{\mbox{\scriptsize p}}$) and the payoff of disregarding protection ($P_{\mbox{\scriptsize np}}$) are:
\bea
P_{\mbox{\scriptsize p}}(t) &=& -c-T \frac{\Ip(t)}{\Ip(t)+\Sp(t)} \label{eq:Pp} \\
P_{\mbox{\scriptsize np}}(t) &=& -T  \frac{\Inp(t)}{\Inp(t)+\Snp(t)}.\label{eq:Pnp}
\eea
The parameter $c$ refers to the cost associated to taking the protective measures. This cost may refer to the monetary cost that the individual has to assume for adopting the protection, but also to other related costs, like the side effects provoked by the protection, or other personal concerns regarding the measures to be taken. The parameter $T$ accounts for the cost of contracting an infection, that is, how severe are the consequences of an infection. The quantities $\frac{\Ip(t)}{\Ip(t)+\Sp(t)}$ and $\frac{\Inp(t)}{\Inp(t)+\Snp(t)}$ are the fraction of infected individuals that have chosen strategy P or NP, respectively. These ratios can be interpreted as the success of one strategy over the other, and players in P and NP use this information to assess how well their strategy is paying off. Therefore, the severity (or cost) of the infection $T$ multiplied by the estimated fraction of protected or not protected individuals that get infected, encapsulates the perceived risk of infection.

The transition probabilities between strategies are defined as a function of the difference in payoffs $\Delta P_{\mbox{\scriptsize np-p}}(t) = P_{\mbox{\scriptsize np}}(t) - P_{\mbox{\scriptsize p}}(t)$ and $\Delta P_{\mbox{\scriptsize p-np}}(t) = P_{\mbox{\scriptsize p}}(t) - P_{\mbox{\scriptsize np}}(t)$, respectively. In this sense, agents will transition to the strategy which is providing a greater payoff at the current time with a given probability. These transition probabilities are those leading to the  replicator dynamics at the population level~\cite{Gintis}, i.e.:
\bea
\Tpnp(t) = \frac{\Delta P_{\mbox{\scriptsize np-p}}(t)}{T+c} \Theta(\Delta P_{\mbox{\scriptsize np-p}}(t)) \label{eq:Tpnp} \\
\Tnpp(t) = \frac{\Delta P_{\mbox{\scriptsize p-np}}(t)}{T+c} \Theta(\Delta P_{\mbox{\scriptsize p-np}}(t))\label{eq:Tnpp},
\eea
\noindent with $\Theta$ representing the Heaviside function, where $\Theta(x)=1$ if $x \geq 0$ and $\Theta(x)=0$ if $x < 0$, and $(T+c)$ being the normalizing factor that is equal to the maximum possible payoff difference between P and NP strategists. 

In general, behavioral changes towards protection do not imply an absolute protection from the disease (e.g.\ the efficacy of condoms for the transmission of HIV is approximately 80\%~\cite{Weller02}). For this reason, we define a parameter $\gamma$ as the probability of the preventive measures failing, where $\gamma=1$ means that the prevention strategy is useless and both protected and unprotected users will get infected with the same probability. On the other hand, if $\gamma=0$, a susceptible protected agent will be totally immune against getting infected, and an infected protected agent will absolutely not transmit the disease to anyone else. The protection mechanism is bilateral, meaning that as long as one of the two parties participating in an infection contact is protected, the other party is protected as well. 
%This is reflected by the linear reduction, instead of quadratic, of the $\gamma$ parameter as two agents with the same P strategy meet (see Eq.~\eqref{eq_react_pp}). The fact that the reduction is linear implies that there is no additional benefit in both partners being protected. This is true for certain protection mechanisms, while for others a quadratic reduction would be in order. For the sake of simplicity, we have considered a linear reduction on $\gamma$, although considering a quadratic reduction would not qualitatively alter the results. 
%(see Methods: Protection threshold). 

The transition probabilities for changing the epidemic compartment can be summarized as the following reactions between agents $i$ and $j$: 
\bea
\Snp^i + \Inp^j & \xrightarrow{\lambda} & \Inp^i+\Inp^j \\
\Snp^i + \Ip^j & \xrightarrow{\gamma\lambda} &  \Inp^i +  \Ip^j\\
\Sp^i + \Inp^j & \xrightarrow{\gamma\lambda} & \Ip^i +  \Inp^j \\
\Sp^i + \Ip^j & \xrightarrow{\gamma\lambda} & \Ip^i + \Ip^j, \label{eq_react_pp}
\eea
\noindent where $\lambda$ is the infectivity rate of the epidemic and $\gamma$ is the aforementioned probability that preventive measures fail. The variables $\Sp^i, \Ip^i, \Snp^i, \Inp^i$ with $i = 1,2, \dots ,N$ describe the state of the $N$ agents in the population. Note that  the linear reduction, instead of quadratic, of the $\gamma$ parameter as two agents with the same P strategy meet (see Eq.~\eqref{eq_react_pp}). The fact that the reduction of the infectivity rate $\lambda$ is linear in $\gamma$ implies that there is no additional benefit in both partners being protected. This is true for certain protection mechanisms, while for others a quadratic reduction would be in order. 
%For the sake of simplicity, we have considered a linear reduction on $\gamma$, although considering a quadratic reduction would not qualitatively alter the results. 

Now, by using all of the above definitions, we can write the dynamical equations of the coupled the risk-driven epidemic model using a probabilistic approach, the Microscopic Markov Chain Approach~\cite{Gomez2010}:
\begin{widetext}
\bea
	\Sp^i(t+1) &=& (1-\Tpnp(t) )\left[ \Sp^i(t)(1-\qp^i(t))+\Ip^i(t) \mu \right]+\Tnpp(t) \left[ \Snp^i(t)(1-\qp^i(t))+\Inp^i(t) \mu \right] \label{eq:Sp}\\
	\Snp^i(t+1) &=& \Tpnp(t) \left[ \Sp^i(t)(1-\qnp^i(t))+\Ip^i(t) \mu \right]+(1-\Tnpp(t) )\left[ \Snp^i(t)(1-\qnp^i(t))+\Inp^i(t) \mu \right] \label{eq:Snp}\\
	\Ip^i(t+1) &=& (1-\Tpnp(t) )\left[ \Sp^i(t) \qp^i(t)+\Ip^i(t) (1-\mu) \right]+\Tnpp(t) \left[ \Snp^i(t) \qp^i(t)+\Inp^i(t) (1-\mu) \right] \label{eq:Ip}\\
	\Inp^i(t+1) &=& \Tpnp(t) \left[ \Sp^i(t) \qnp^i(t)+\Ip^i(t) (1-\mu) \right]+(1-\Tnpp(t) )\left[ \Snp^i(t) \qnp^i(t)+\Inp^i(t) (1-\mu) \right] \label{eq:Inp}
\eea 
\end{widetext}
\noindent where $\mu$ is the epidemic recovery rate. The terms in brackets in the equations refer to the epidemic spreading dynamics, which describe the transit between the compartments $S \rightleftharpoons I$. The other terms refer to the game dynamics, allowing the transition between the compartments $P \rightleftharpoons NP$. The quantities $\qp^i(t)$ and $\qnp^i(t)$ express the probability that agent $i$ will get infected at time $t$ if she is protected or not protected, respectively, and read:
\bea
\label{eq:qvqnv}
\qp^i (t) & = 1-\prod\limits_{j=1}^{N} \left[ 1-A_{ij}\lambda \gamma (\Ip^j(t)+\Inp^j(t)) \right]  \\
\qnp^i(t) & = 1-\prod\limits_{j=1}^{N} \left[ 1-A_{ij}\lambda (\gamma \Ip^j(t)+\Inp^j(t)) \right],
\eea 
\noindent where $A_{ij}$ refers to the adjacency matrix of the epidemic process, with $A_{ij}=1$ if agents $i$ and $j$ are connected, and $A_{ij}=0$ otherwise. We are implicitly assuming a markovian dynamics, i.e. that temporal correlations are absent on active edges~\cite{lai19}.

%%-------------------------------------------------------%%
%%          OSCILLATIONS                              %%   % < 527 words
%%-------------------------------------------------------%%
\section{Oscillatory behavior} 
\begin{figure}[ht]
	\centering
	\includegraphics[width=1.00\columnwidth]{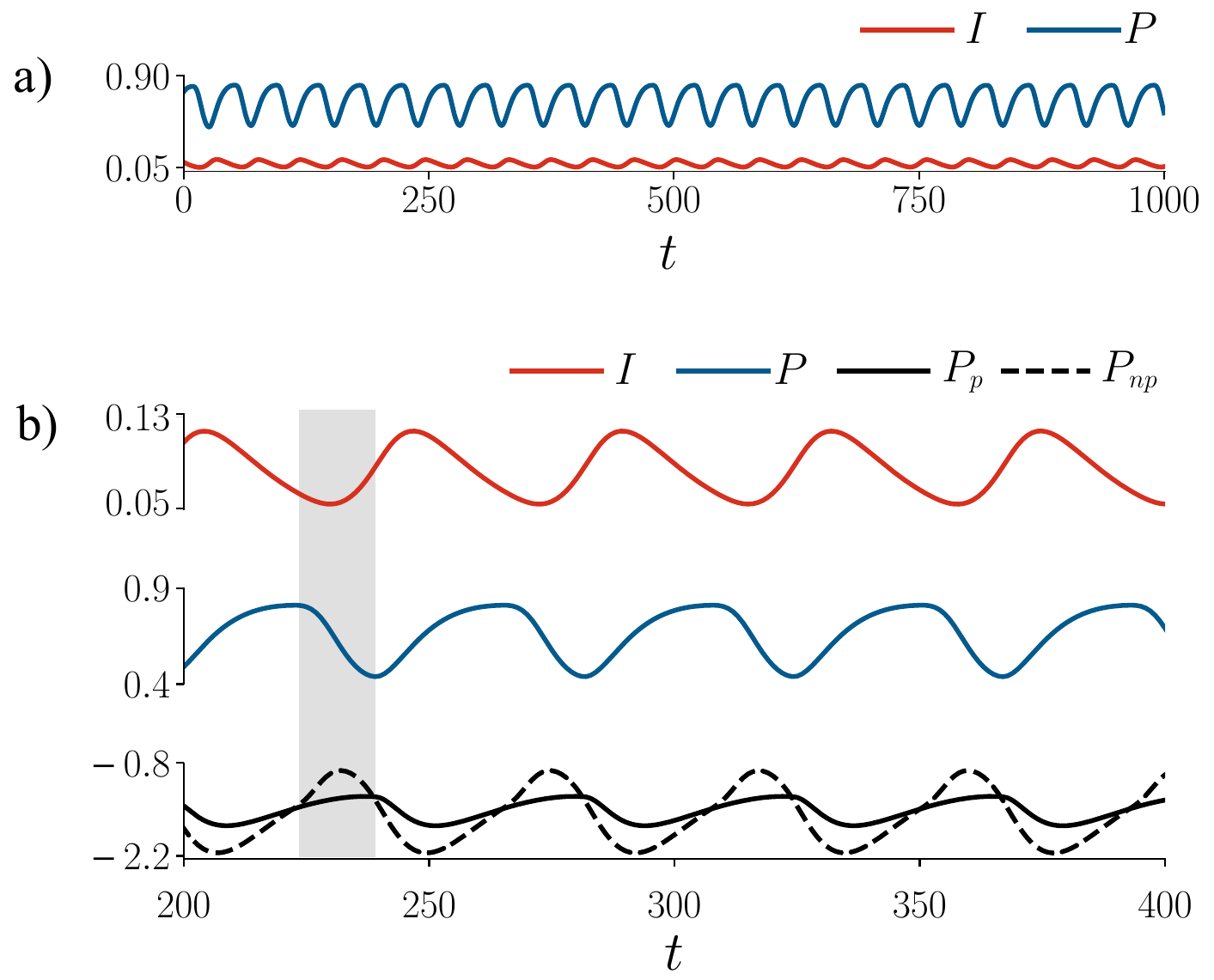}
	\caption{\footnotesize 
	%{Risk-driven epidemic spreading model.} 
	Numerical results of the risk--driven epidemic spreading model on a power-law network of size $N=2000$ and exponent $2.5$. Default parameters are $c=1$, $\mu=0.1$, $T=10$, $\lambda=0.05$, and $\gamma=0.1$. 
	\textbf{(a)} Fraction of Protected ($P=\Sp+\Ip$) and Infected ($I=\Ip+\Inp$) individuals as a function of time. We observe an oscillatory behavior that is sustained in time. 
	\textbf{(b)} Detail of the oscillations. The red and blue lines indicate the fraction of Infected and Protected individuals, respectively. The black dashed line plots the payoff of the strategy {\em not protected} ($\Pnp$) while the solid black line is the payoff of the {\em protected} strategy ($Pp$).}
	\label{fig:figure1}
\end{figure}

Solving numerically the equations of the risk--driven epidemic model (Eqs.~\eqref{eq:Sp}-\eqref{eq:Inp}) on top of network models (see Appendix A), we observe that the infection prevalence, $I$, as well as the number of protected individuals, $P$, oscillates in time in a sustained way, see Fig.~\ref{fig:figure1}(a). To unveil the mechanism behind the oscillations, we plot in Fig.~\ref{fig:figure1}(b) the fraction of agents in each one of the compartments, in time. If we focus on the gray area, we see that when $\Pnp$ is higher than $\Pp$, individuals cease to protect themselves, and this implies that the number of infected individuals start to increase. When this happens, the payoff of the strategy {\em protected} ($\Pp$) becomes larger than $\Pnp$, provoking individuals to start protecting themselves, and thus the number of infected individuals is again reduced, sustaining the limit cycle observed in the evolution of $I$ and $P$.

% ____ NEW FIGURE 1  ________

An interesting question is whether oscillations disappear or are reduced for certain values of the parameters. We explored how the cost of contracting the disease $T$ affects the aforementioned oscillations, and observed that higher values of $T$ (higher cost) generate smaller oscillations (see Fig.~\ref{fig:figure1}(a) left plot), given that when the cost of contracting an infection is really high, almost all individuals adopt the {\em protected} strategy. However, when plotting the relative amplitude of the oscillations (see Fig.~\ref{fig:figure1}(a)(right)) we observe that for all values of $T$ the relative amplitude has the same order of magnitude. The absolute value of the oscillations is lowered due to a lower presence of infectious cases, but relatively the oscillations are the same. We conclude that the cost of contracting the disease $T$ does not induce the oscillations to vanish.

% ____ FIGURE 2 ________
\begin{figure}[htbp]
	\centering
	\includegraphics[width=1.00\columnwidth]{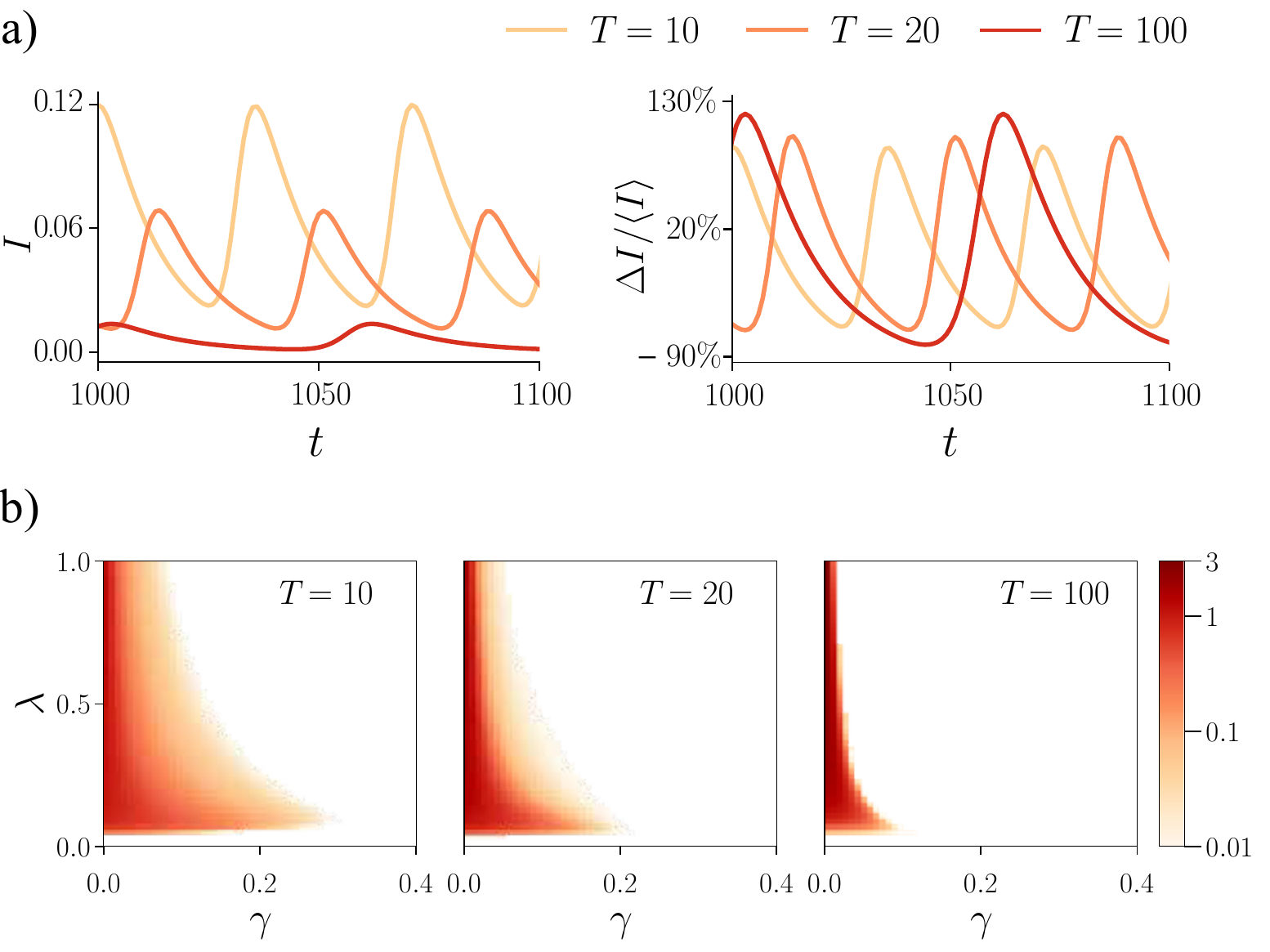}
	\caption{\footnotesize 
	%{Amplitude of the oscillations.} 
	Numerical results of the risk--driven epidemic spreading model on a power-law network of size $N=2000$ and exponent $2.5$. Default parameters are $c=1$, $\mu=0.1$, $T=10$, $\lambda=0.05$, and $\gamma=0.1$. 
	\textbf{(a)} Amplitudes of the oscillations in the fraction of infected individuals as a function of time. Left plot depicts the absolute value of the amplitude, while the right one depicts the relative one, all of them for three values of $T$, the cost of infection. Higher values of $T$ (higher cost) generate smaller oscillations (see left plot). On the right, we can see that all amplitude of oscillations are of the same order of magnitude, when calculated relatively to the fraction of infected individuals. 
	\textbf{(b)} Average relative amplitudes of the fraction of infected individuals in the steady state, for all range of $\gamma$ (the probability of protection failure) and $\lambda$ (the infectivity rate), for different values of $T$. We observe that as the infection cost increases, the area where the oscillations are present is reduced, but the oscillations themselves are larger.}
	\label{fig:figure2}
\end{figure}

One would think that the probability that preventive measures fail ($\gamma$) and the infectivity rate ($\lambda$) are able to shape the oscillations as well. We explore this in Fig.~\ref{fig:figure2}(b), and find that oscillations are only present for low values of $\gamma$ and low values of $\lambda$, pointing out that only when the preventive measure is very effective and the disease is not very contagious is when individuals face the dilemma on whether to protect themselves that ultimately leads to the aforementioned oscillatory behavior. Outside this area of the parameters, either the disease is very contagious, or the measures are useless, or a combination of both. In any case, the number of infected individuals grows larger and we do not observe oscillations. We also see that as the infection cost $T$ increases, the region of parameters that presents oscillations becomes smaller, but the relative oscillations are higher. 

Note that oscillations can only emerge when the time scale for the decision on prophylaxis is equal or faster than the typical time scale of the disease. Otherwise, the epidemics will reach equilibrium too soon for any strategic decision to have impact on the infection. On the other hand, when agents evaluate their payoffs before the disease has reached its equilibrium, the success of each strategy (used in  Eqs.~\eqref{eq:Pp} and \eqref{eq:Pnp} to update the prophylactic behavior) does not capture the actual risk of getting infected. In other words, agents take their decisions based on information that is delayed in time. This is precisely the most common scenario for many epidemic outbreaks.

\begin{figure}[htbp]
	\centering
	\includegraphics[width=0.90\columnwidth]{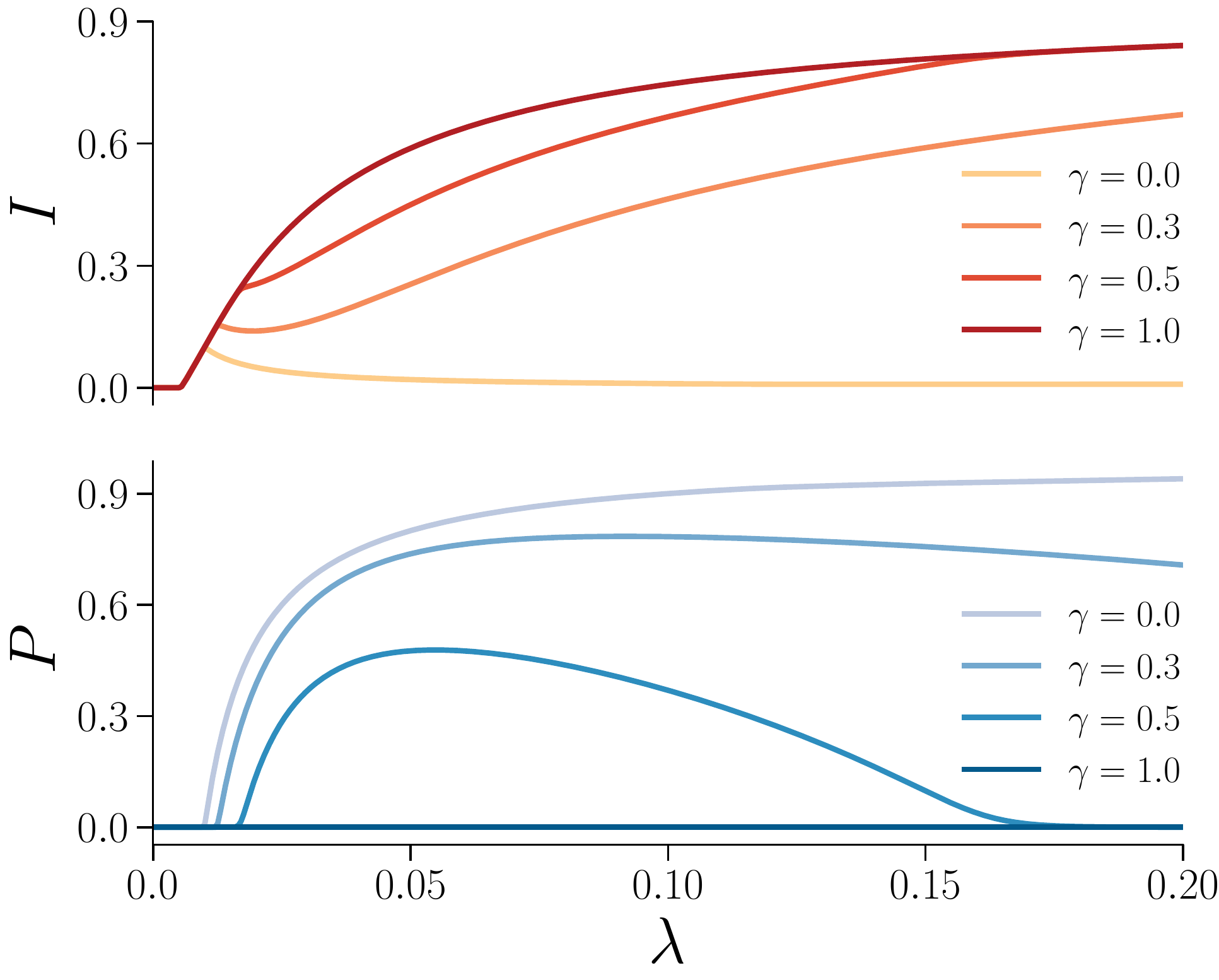}
	\caption{\footnotesize Numerical results of the proposed model on a power-law network of size $N=2000$  and exponent $2.5$. Default parameters are $c=1$, $\mu=0.1$, and $T=10$. Fraction of infected individuals ($I=\Inp+\Ip$) and fraction of protected individuals ($P=\Sp+\Ip$) in the steady state as a function of the epidemic infectivity probability $\lambda$, for different values of the probability of preventive measures failing ($\gamma$).}
	\label{fig:figure3}
\end{figure}

%%--------------------------------------------------------------------%%
%%          Effectiveness of the protection method           %%  % Words = approx < 800w
%%---------------------------------------------------------------------%%
%\section{Effectiveness of the protection method}
Once explored the nature of the oscillatory dynamics, we now characterize to what extent the quality of the protective measures affects the spreading of the disease. In Fig.~\ref{fig:figure3} we plot the steady state fraction of both $I$ and $P$ individuals as a function of the epidemic infectivity $\lambda$, for different protection effectivity values $\gamma$. When $\gamma=1$, the prevention measures are useless and thus we recover the second order phase transition typical of an SIS model. For very effective measures ($\gamma \approx 0$), a majority of the population adopts the protective behavior and thus the number of infected individuals is almost zero for all values of $\lambda$. For intermediate values of the protection effectivity ($\gamma=0.5$) the effect is interesting: for values of the infectivity sufficiently low (but above the critical threshold), a fraction of individuals adopt protection, but as the infectivity increases, for this particular value of the effectivity $\gamma$, prophylaxis is not enough to prevent infection and thus agents cease to protect themselves. Looking at the top plot, we see that the adoption of protection in this region of $\lambda$ leads to a decrease in the number of infections, but the range of $\lambda$ where this happens is small. In conclusion, the addition of prophylaxis is only able to diminish (not completely eliminate) the number of infectious cases and that only happens for a narrow range of $\lambda$. Actually, the epidemic is only fully eradicated when the disease infectivity $\lambda$ is below its critical value $\lambda_c$, and this critical value does not seem to depend on the decision game. 

% _____ FIGURE 3  _______

\section{Epidemic threshold}

To see if the latter statement is true, we need to calculate the critical threshold of this coupled system, and discern which are the parameters that influence it. 
We start off by considering our system in equilibrium, i.e. the game does not evolve and the epidemics is stationary. The equilibrium of the decision game is reached when agents keep the same strategy over time. For an agent to keep the same strategy over a period of time, it must happen that no other strategy provides a better payoff than the one provided by the current strategy. In our case, this only happens when the transition probabilities between strategies fulfil $\Tnpp = \Tpnp = 0$, which in turn requires that both payoffs are equal, i.e. $\Pp = \Pnp$ (see its derivation in Appendix B). Using Eqs.~\eqref{eq:Pp}-\eqref{eq:Pnp} we have:
\beq
\label{eq:eqGame}
\frac{c}{T} = \frac{\Inp}{\Inp+\Snp}-\frac{\Ip}{\Ip+\Sp}.
\eeq
However, imposing the previous condition for the equilibrium of the game at the critical point of the epidemics $\lambda_c$, and noting that at $\lambda_c$ the fraction of $\Ip,\Inp \approx 0$, we observe that this condition can only be satisfied for $c=0$. In other words, the equilibrium of the system at the critical point can only be achieved when the cost of the protection is zero, which gives no incentives to individuals to choose the protected strategy over the non--protected one, according to Eqs.~\eqref{eq:Pp}-\eqref{eq:Pnp}. 	

In conclusion, at the critical point of the epidemics the game plays no role, and thus the epidemic threshold is the usual of the SIS model ~\cite{Gomez2010}:
\beq
\lambda_c = \mu / \Lambda_{\text{max}} (A), \label{eq:epiThreshold}
\eeq
\noindent where $\Lambda_{\text{max}} (A)$ refers to the maximum eigenvalue of the adjacency matrix $A$. 
%
%The calculation of the epidemic threshold reveals that at the critical point of the epidemics the game has no effect, and thus we obtain the well-known epidemic threshold for the SIS model:
%\beq
%\lambda_c = \mu / \Lambda_{\text{max}} (A), \label{eq:epiThreshold}
%\eeq
%\noindent where $\Lambda_{\text{max}} (A)$ refers to the maximum eigenvalue of the adjacency matrix $A$. 
Indeed, this result confirms our intuition that the decision game is not able to shift the tipping point of the epidemics.

\section{Protection threshold}
In analogy to the epidemic threshold, we can define the threshold $\tilde{\lambda}$, such that agents start protecting themselves. The protection threshold $\tilde{\lambda}$, can be calculated in a well-mixed population assuming that the fraction of protected agents is negligible in comparison to the fraction of non-protected ones. In the well mixed population, agents interact randomly. Accordingly, the interactions are not structured and thus the probability for an agent to be in a given compartment is the same in the whole population. Therefore, the system can be described by only four variables: $\Sp$, $\Snp$, $\Ip$, $\Inp$. Accordingly, the recurrence relations take the same form as in Eqs.~\eqref{eq:Sp}-\eqref{eq:Inp} but without the label $i$. In a similar way, the infection probabilities $\qp$ and $\qnp$ are given by:
	\bea\nonumber
	\qnp(t) = \lambda (\gamma \Ip(t) + \Inp(t))\\
	\qp(t) = \lambda \gamma (\Ip(t) + \Inp(t)) .
	\eea
	For calculating this threshold in a well--mixed population, $\tilde{\lambda}$, we can make use of the equilibrium condition (that translates in a stationary state for the epidemics). 
	We recall that the protection threshold will satisfy: $\Sp,\Ip \ll \Snp,\Inp$ and $\Sp,\Ip \ll 1$. Therefore, we can expand the equilibrium condition for the game in Eq.~\eqref{eq:eqGame} as:
	\bea
	\label{eq:threshold0}\nonumber
	\frac{c}{T}(\Ip&+&\Sp ) = \frac{\Inp}{(\Ip +\Sp )}{1-\Ip -\Sp }-\Ip \\
	&=& \Inp(\Ip+\Sp)-\Ip+\mathcal{O}((\Ip +\Sp)^{2}).
	\eea
Since we want to calculate the threshold for the strategy $p$, we will neglect all the second order and higher order terms in $(\Ip +\Sp)$. Consequently, Eq.~\eqref{eq:threshold0} can be rewritten as:
\beq
	\label{eq:threshold1}
	\Ip = (\Ip+\Sp)\left( \Inp-\frac{c}{T}\right)\,,
	\eeq
	
	\noindent  and the stationary Eqs.~\eqref{eq:Sp}-\eqref{eq:Inp}, in the well-mixed approximation, read:
	%neglecting the second order terms is equivalent to neglecting the risk that a susceptible agent with strategy $p$, $\Sp$,  gets infected by another agent with strategy $p$, $\Ip$. 
	%Therefore, whether the transmission probability is reduced linearly or quadratically as two agents with strategy $p$ meet does not influence the threshold. 
	%Consequently, the here presented calculation of the threshold would also be valid if we assumed that agents with strategy $p$ reduce their interaction rate. 
	%neglecting the second order terms, the equilibrium conditions become:
\bea
	\Sp &=& \frac{\Ip}{\Inp}\frac{\mu}{\gamma \lambda} \label{eq:threshold2}\\ 
	\Ip &=& \Inp \frac{1-\Inp-\frac{\mu}{\lambda}}{\Inp(1+\gamma)-\gamma+\frac{\mu}{\gamma \lambda}}. \label{eq:threshold3}
	\eea
Inserting Eq.~\eqref{eq:threshold2} into Eq.~\eqref{eq:threshold1} leads to a quadratic equation for $\Inp$:
\beq
	\label{eq:threshold4}
	\Inp^{2} - \Inp\left(1+\frac{c}{T}-\frac{\mu}{\gamma \lambda}\right)-\frac{c}{T}\frac{\mu}{\gamma \lambda} = 0\,.
	\eeq
In the quadratic equation, there is only one non negative solution given by:
\beq
	\label{eq:threshold5}
	\Inp = \frac{1}{2}\left[1-\frac{\mu}{\gamma \lambda}+\frac{c}{T}+\sqrt{\underbrace{\left(\frac{\mu}{\gamma \lambda}+\frac{c}{T}\right)^{2} +4\frac{\mu}{\gamma \lambda}}_{\Delta}}\right]\,.
	\eeq
	
	%\noindent The threshold can now be found if we consider the condition such that $P \neq 0$. Considering Eq.~\eqref{eq:threshold2}, we have $\Sp > 0 \iff \Ip > 0$. Therefore, 
\noindent The threshold, $\tilde{\lambda}$, is reached as $\Ip$ becomes non zero. The denominator in the expression of $\Ip$ in Eq.~\eqref{eq:threshold2} is always positive: 	
%	\bea\nonumber
%	\Inp(1+\gamma)-\gamma+\frac{\mu}{\gamma \lambda} &=& \frac{1}{2}\left[ 1-\gamma+\frac{\mu}{\lambda}\left(\frac{1}{\gamma}-1\right)\\
%	&+&(1+\gamma)\left(\frac{c}{T}+\sqrt{\Delta}\right)\right] > 0\,.
%	\eea
	\begin{multline}
	\Inp(1+\gamma)-\gamma+\frac{\mu}{\gamma \lambda} = \\
	\frac{1}{2}\left[ 1-\gamma+\frac{\mu}{\lambda}\left(\frac{1}{\gamma}-1\right)+(1+\gamma)\left(\frac{c}{T}+\sqrt{\Delta}\right)\right] > 0\,.
	%\end{equation}
	\end{multline}
The positivity is guaranteed due to $\gamma \in [0,1]$. Since the denominator is always positive, the threshold is then reached when the numerator becomes zero. The threshold condition therefore reads: 
\beq
1-\Inp-\mu / \tilde{\lambda} = 0. \label{eq:cond}
\eeq
Inserting the expression of $\Inp$ in Eq.~\eqref{eq:cond}, we find, after some algebra, the threshold $\tilde{\lambda}$:
\beq
\label{eq:protectionThreshold}
\tilde{\lambda}^{\pm} = \frac{2\mu}{1-\frac{c}{T}\mp \sqrt{\left(1-\frac{c}{T}\right)^2-4\frac{c}{T}\frac{\gamma}{1-\gamma}}}.
\eeq

%, as presented in Eq.~\eqref{eq:protectionThreshold}. In order to get a more accurate description of the networked dynamics, we can consider a well mixed population in which agents interact $k$ times at each time step. The number of interactions $k$ is then fixed by the mean degree of the network we want to approximate. the protection threshold is then rescaled as $\tilde{\lambda} \rightarrow \tilde{\lambda}k$.
%
%
%
%
% (see Methods: Protection threshold), and we obtain:
\noindent The equation for the threshold has two solutions: $\tilde{\lambda}^{-} $ and $\tilde{\lambda}^{+} $. The threshold $\tilde{\lambda}^{-}$ describes the critical infectivity rate above which agents start protecting themselves. In other words, below $\tilde{\lambda}^{-}$ there is still no risk of infection sufficiently high as to consider taking preventive measures. On the other hand, the threshold $\tilde{\lambda}^{+}$ is the point where agents stop adopting the protective behavior, since the protection is not sufficient to combat the high infection risk. From Eq.~\eqref{eq:protectionThreshold} we obtain $\lim_{\gamma \rightarrow 0} \tilde{\lambda}^{+} = \infty$. This means that, under the condition where protection leads to complete immunization ($\gamma=0$), an increasing infectivity rate does not stop agents from taking preventive measures against the disease and the disease is controlled independently of the infection probability. 

% _____ FIGURE 4  _______
\begin{figure}[htbp]
	\centering
	\includegraphics[width=1.00\columnwidth]{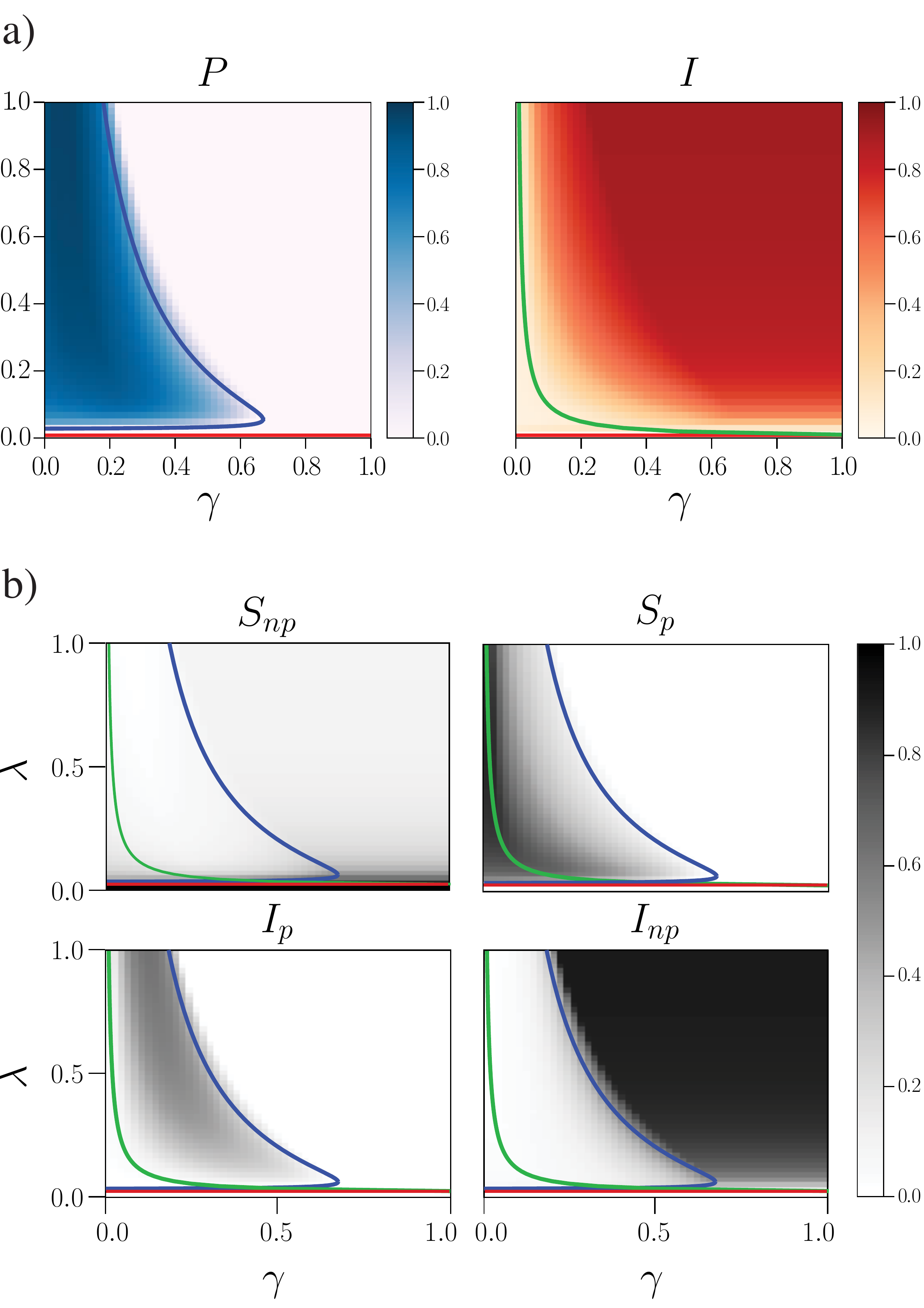}
	\caption{\footnotesize Numerical results of the proposed model on a power-law network of size $N=2000$  and exponent $2.5$. Default parameters are $c=1$, $\mu=0.1$, and $T=10$. \textbf{(a)} Phase-space diagrams of the prevalence on the number of Protected (left) and number of Infected individuals (right). Red line in left plot denotes the epidemic threshold of our model, as calculated by \eqref{eq:epiThreshold}. The blue line on the left plot is the protection threshold as obtained in \eqref{eq:protectionThreshold}. The green line indicates the epidemic threshold in the case of a fully protected population. \textbf{(b)} Phase-space diagrams for each one of the four compartments of our model, at the steady state, for all range of $\lambda$ and $\gamma$. }
	\label{fig:figure4}
\end{figure}

In Fig.~\ref{fig:figure4}(a)(left) and Fig.~\ref{fig:figure4}(a)(right), we present the full phase space ($\lambda-\gamma$) for the expected fraction of protected individuals, and infected individuals, respectively. We see clearly that agents only take a protective behavior when the infectivity rate is low and the preventive measures are reasonably efficient. The real parts of both solutions $\tilde{\lambda}^{-} $ and $\tilde{\lambda}^{+} $ are displayed in Fig.~\ref{fig:figure4}(a)(left) as a single blue curve over the phase space of the prevalence of the protected.

To further understand the adoption of prophylactic behavior, in Fig.~\ref{fig:figure4}(b) we show the partition of the population across the four compartments ($\Snp$, $\Sp$, $\Ip$ and $\Inp$) as a function of $\lambda$ and $\gamma$. From the analysis of the four compartments we observe four different regimes: 
\begin{enumerate}[label=\Roman*.]
\item Below the epidemic threshold $\lambda_c=0.01$ (red line), the disease dies out and thus all agents are in the $\Snp$ compartment.
\item For low values of $\gamma$ (meaning high protection effectivity), the majority of agents adopt the protection and avoid the infection, this is observed at the darkest area of the $\Sp$ compartment (below the green line that corresponds to a fully protected population).
\item For values of $\gamma$ in the protection range defined by the thresholds Eq.~\eqref{eq:protectionThreshold} (area delimited by the blue line) most agents protect themselves but still get infected.
\item Beyond the protective threshold (area beyond the blue line), the protection measures are highly ineffective, which causes agents to disregard any protection and thus the infection prevalence is the one expected for such values of $\lambda$ in absence of the decision game.
\end{enumerate}
%(I) Below the epidemic threshold $\lambda_c=0.01$ (red line), the disease dies out and thus all agents are in the $\Snp$ compartment. (II) For low values of $\gamma$ (meaning high protection effectivity), the majority of agents adopt the protection and avoid the infection, this is observed at the darkest area of the $\Sp$ compartment (below the green line that corresponds to a fully protected population). (III) For values of $\gamma$ in the protection range defined by the thresholds Eq.~\eqref{eq:protectionThreshold} (area delimited by the blue line) most agents protect themselves but still get infected. (IV) Beyond the protective threshold (area beyond the blue line), the protection measures are highly ineffective, which causes agents to disregard any protection and thus the infection prevalence is the one expected for such values of $\lambda$ in absence of the decision game.

%%-------------------------------------------------------%%
%%          Intervention policies                        %%    % Words = approx < 1030
%%-------------------------------------------------------%%
\section{Intervention policies}
Once we understand the key role that risk perception has on the adoption of protective measures and, in turn, on the infection prevalence, we now focus on possible intervention strategies. Such strategies are often designed to change the individual's risk perception, as this is known to induce a behavioral change~\cite{kasperson1998,Bagnoli2007}. One way to change the risk perception is to spread information about the severity of the disease in the hopes of raising awareness and containing the epidemic spreading. This can be done either locally, by considering first-hand information and word-of-mouth spreading~\cite{funk2009,Perra2011,Granell2013,Massaro2014}, or globally, i.e. using mass-media outlets to disseminate such information~\cite{Granell2014}. Our proposal is to raise awareness globally, by increasing the perception of risk of the population. In our model, the risk perception is encapsulated by the cost of the infection $T$ multiplied by the estimated fraction of protected or not protected individuals that get infected, respectively. Our strategy relies on increasing the cost of infection $T$ by some quantity $\Delta T$, which can be interpreted as tricking the population into believing that the consequences of an infection are more severe than they actually are. Based on the previous mechanism, we propose two types of campaigns: (i) awareness campaigns continuously enforced in time, and (ii) pulsating awareness campaigns only activated when the infection prevalence is increasing.

In the first one, the campaign raises the perception risk continuously by a certain increment $\Delta T$. Given that this $\Delta T$ is sustained in time, this simply leads to an increased infection risk $(T + \Delta T)$ in Eqs.~\eqref{eq:Pp}-\eqref{eq:Pnp}. 
In the second one, the increment $\Delta T$ is applied only when the infection prevalence is growing with time, {\em i.e.} $(T + \Delta T \Theta[I(t)-I(t-1)])$, where $\Theta(x)$ is the Heaviside function. Additionally, for the pulsed intervention we normalize the transition probabilities in Eqs.~\eqref{eq:Tpnp}-\eqref{eq:Tnpp} according to the infection cost at time $t$, \ie with $T+c+\Delta T \Theta [I(t)-I(t-1)]$.

\begin{figure}[htbp]
	\centering
	\includegraphics[width=1.00\columnwidth]{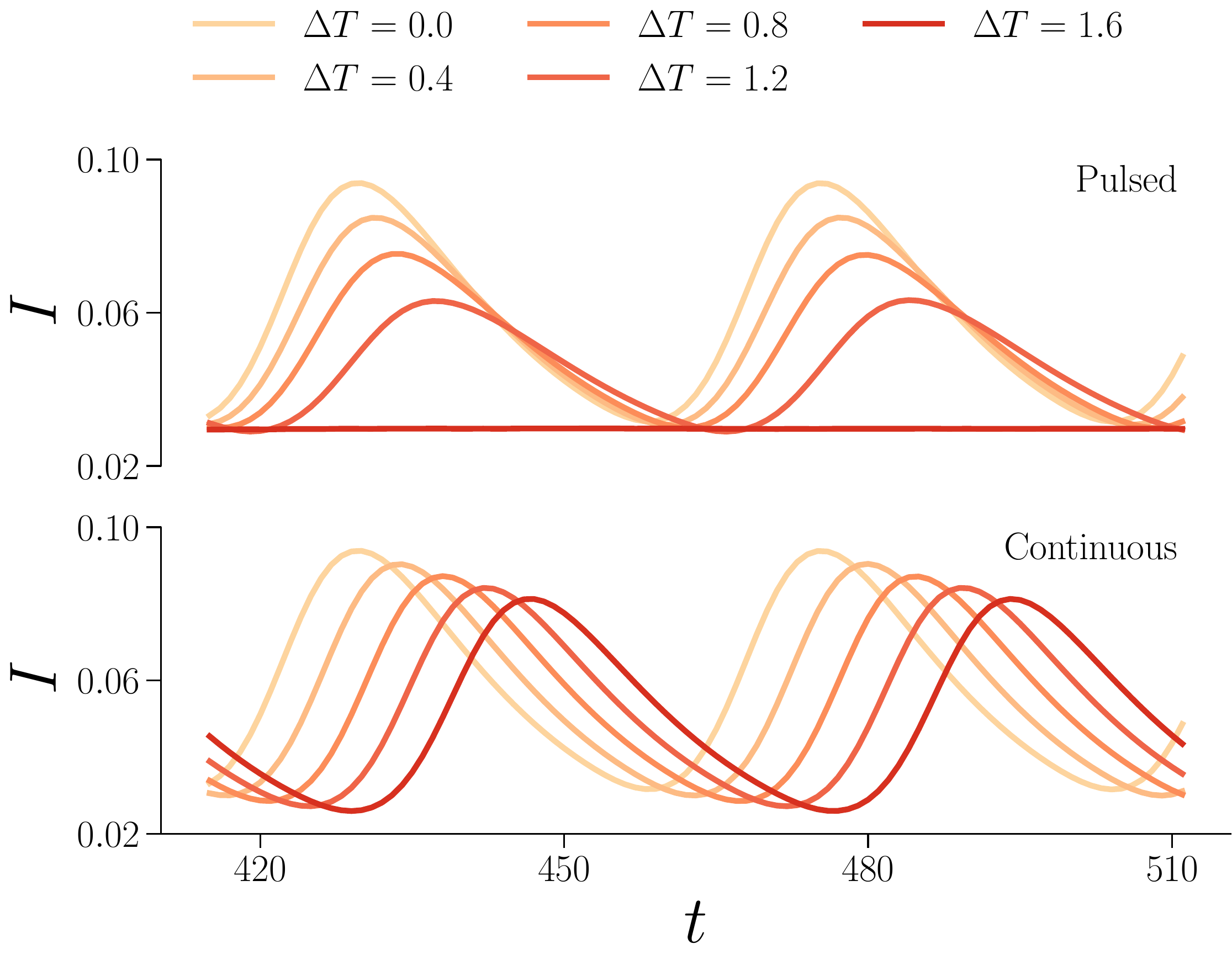}
	\caption{Time evolution of the infection prevalence in the presence of a pulsating awareness campaign (top) and  a continuously sustained awareness campaign (bottom), for different values of the perceived risk increment $\Delta T$. The contact network used here is a power-law network of size $N=2000$  and exponent $2.5$.}
	\label{fig:figure5}
\end{figure}

\begin{figure}[htbp]
	\centering
	\includegraphics[width=1.00\columnwidth]{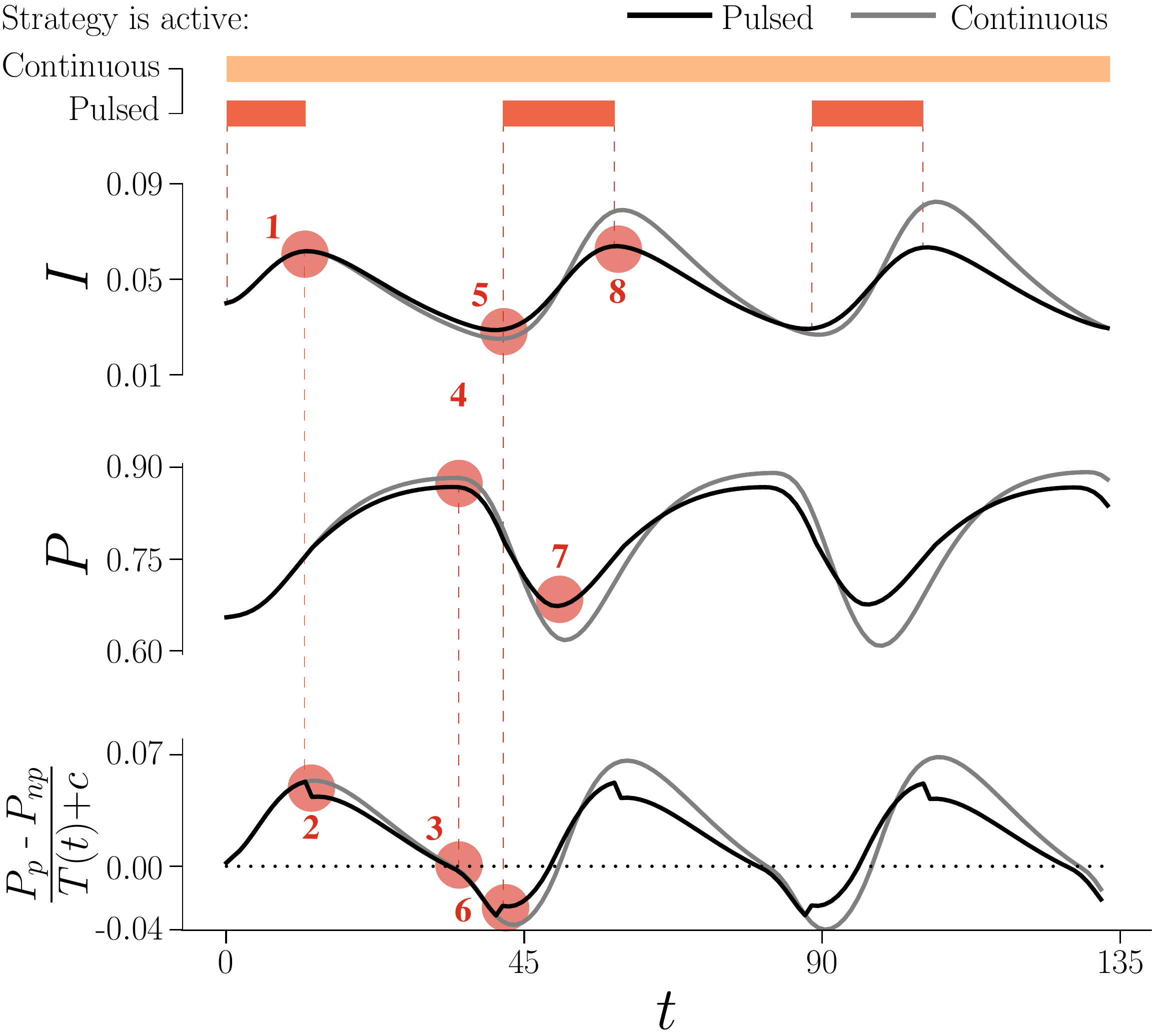}
	\caption{Infection prevalence (top), protection level (middle) and normalized payoff difference (bottom), as a function of time for $\Delta T = 1.2$. For the pulsed intervention, we define $T(t) = T+\Delta T \Theta [I(t)-I(t-1)]$. In the case of a continuous intervention, we have $T(t) = T + \Delta T$ . The contact network used here is a power-law network of size $N=2000$  and exponent $2.5$. Default parameters are $c=1$, $\mu=0.1$, and $T=10$. The pulsating campaign allows to more efficiently suppress the peaks in the infection prevalence than the sustained campaign.}
	\label{fig:figure6}
\end{figure}

Both campaigns increase the population's perception on how serious it is to contract an infection, the only difference being whether this intervention is enforced temporarily (in the case of pulsating) or permanently (in the continuous case). What we observe is that this strategy turns out to be more effective in alleviating the oscillations when it is enforced in a pulsating manner than when it is promoted continuously, see Fig.~\ref{fig:figure5}. This result seems counterintuitive at first, as one would think that a constant and permanent shift in the infectivity cost would be more effective than a shift that is only applied on and off. To illustrate the mechanism why a pulsating campaign proves more effective than a continuous one, we plot, in Fig.~\ref{fig:figure6}, the fraction of infected individuals {\em I}, the protection level {\em P}, and the normalized payoff difference between the protected and non-protected strategies, as they evolve in time, starting from $t=0$. On top, for the sake of clarity, we illustrate the periods of time in which each of the campaigns is active.

% _____ FIGURE 5 : Intervention campaigns: time evolution ______

% _____ FIGURE 6: Intervention campaigns: sketch of mechanism ______

Following the enumerated points depicted in Fig.~\ref{fig:figure6}, we are able to describe the effective differences between the two strategies. Starting out from an initial fraction of infected individuals, the infection starts to grow. In this stage, both interventions are active and operate with the same infection cost $T+\Delta T$, therefore rendering identical curves. However, as the infection grows, individuals start protecting themselves, which slows down the spreading of the infection. Eventually, the protection is adopted by enough individuals such as to prevent the infection from increasing any more. This maximum of the infection is depicted in label 1. Given that the epidemics is not increasing anymore, the pulsed strategy is switched off, which causes the difference in payoffs to drop for the pulsed strategy (see label 2). This means that, at this point, individuals subject to the continuous intervention have a greater incentive to adopt protection than those under the pulsating campaign. The normalized payoff difference for both campaigns evolve further, up to a point in which the payoff for the protected strategy and the payoff for the non-protected strategy are equal (the payoff difference is zero, see label 3). This point defines the maximum of the protection curve (see label 4), because from that moment on, individuals will have a negative payoff difference, meaning they will consider a better strategy to disregard protection. This, in turn, promotes the propagation of the epidemics, and therefore the {\em I} curve starts to grow again (see label 5). At this moment, the pulsating strategy is switched on again, causing the pulsating intervention's payoff difference to increase abruptly, effectively reaching a lower minimum than the one of the continuous strategy (see label 6). At this point, users subject to the pulsating intervention have both a higher number of infectives than in the continuous case, and a higher payoff difference; which implies that the individuals under the pulsating campaign will perceive a higher risk and protect themselves more than under the continuous intervention. This can be seen in label 7, where individuals in the pulsating campaign are not disregarding protection as much as continuous intervention users. Consequently, the infection prevalence in the case of a pulsating campaign will grow less than the one subject to continuous intervention, as seen in label 8. From that moment on, the mechanism illustrated in labels 1 to 8 is repeated periodically.

Put in a nutshell, both continuous and pulsating campaigns seek to raise the awareness of the population by increasing the perceived infection cost. The reason why the pulsating campaign is more effective is because the minima and maxima of the function of the difference in payoffs are moderated by switching on and off the intervention campaign. When the maximum difference in payoffs is relaxed by switching off the intervention (as seen in label 2), it causes less people to adopt protection (label 4), which results in a slightly higher infection than in the continuous case (label 5). Symmetrically, when the minimum in payoff difference is increased by switching on the intervention (label 6), less people disregard protection (label 7) and a lower infection prevalence is achieved (label 8). By making the minima and maxima less pronounced, the oscillations are more dampened in the pulsating intervention.

%%-------------------------------------------------------%%
%%          CONCLUSIONS                              %%   % Words = 371
%%-------------------------------------------------------%%
\section{Conclusions} 
Wrapping up, we have confronted the quantitative analysis of prophylactic human behavior in the spread of direct contact transmittable diseases using a mathematical model. The results allow us to better understand some observed oscillatory patterns that could depend on the biological seasonality of viruses and bacteria, and according to our findings, also on the human decision of prophylaxis. 

Our model is the first description, up to date, of the co-evolutionary dynamics of human behavior and disease spreading, using a probabilistic microscopic model for the epidemic spreading coupled to a risk-driven decision game. The accuracy of the epidemic spreading description at the level of individuals in a network, using our model, allows to include a decision strategy in a very natural way for each individual. The results of the model are enlightening. First we prove the emergence of self-sustained oscillations in the prevalence of the disease as a consequence of the interplay between the prevalence and the assessment of risk by individuals. This discovery allows to think about mechanisms to ameliorate the evaluation of risk made by individuals, with the aim of damping out these oscillations that represent a health threat and a possible collapse of health services. 

We do fix our attention on the quantitative evaluation of awareness campaigns as probably the best strategies to modify risk perception. We analyzed two different types of campaigns: continuous (persistent in time), and pulsating (active only in certain periods of time). Through the mathematical model we discover that pulsating campaigns are far more efficient that continuous ones  in damping out the oscillatory behavior of the disease. The explanation of this interesting finding is rooted on the interplay between the risk assessment and the real prevalence: the pulsating campaign triggers awareness only when the prevalence is rising and not when it is descending. This provokes a delay on the prophylaxis action for those individuals who assess the risk based on the current prevalence and not on the awareness, and this small mismatch makes the awareness action more effective when it appears again. 
%From a psychological point of view, we could say that the pulsating campaign acts as a periodic alert that makes it more difficult for individuals to forget about the danger of getting infected.

%%-------------------------------------------------------%%
%%          ACKNOWLEDGEMENTS               %%
%%-------------------------------------------------------%%
\section*{Acknowledgements}
\noindent
B.S. acknowledges financial support from the European Union's Horizon 2020 research and innovation programme under the Marie Sklodowska-Curie grant agreement No. 713679 and from the Universitat Rovira i Virgili (URV). A.A. acknowledges financial support from Spanish MINECO (grant PGC2018-094754-B-C21), Generalitat de Catalunya ICREA Academia, and the James S. McDonnell Foundation grant \#220020325. J.G.G. acknowledges financial support from MINECO (projects FIS2015-71582-C2 and FIS2017-87519-P) and from the Departamento de Industria e Innovaci\'on del Gobierno de Arag\'on y Fondo Social Europeo (FENOL group E-19). C.G. acknowledges financial support from Juan de la Cierva-Formaci\'on (Ministerio de Ciencia, Innovaci\'on y Universidades) and the James S. McDonnell Foundation Postdoctoral Fellowship, grant \#220020457.

\vspace{1cm}
\section*{Appendix}

\subsection{Network model} \label{network_model}
To illustrate the model, we will consider the case of sexually transmitted diseases, without considering the medical details of any particular disease, only the direct propagation mechanism and the associated risk perception. The propagation of sexually transmitted diseases takes place on the so-called sexual contact networks~\cite{Liljeros2001,Liljeros2003,Schneeberger2004}. In these networks, the distribution of the number of sexual partners is heterogeneous, with few individuals having a number of sexual contacts orders of magnitude larger than the average. Previous studies approximated the distribution of the number of sexual contacts, $P(k)$, with a power law $P(k) \sim k^{-\gamma}$, being $k$ the number of sexual contacts. These studies~\cite{Schneeberger2004, GomezGardenes2008} identified scaling exponents between $1.5 < \gamma < 3.5$. More specifically, they state that sexual contact networks have different scaling exponents depending on whether they depict relationships between men, men and women, or among women. For the sake of simplicity, we will focus on men--men sexual contact networks, although the same analysis can be performed on heterosexual networks. We will study the coupled disease-decision dynamics on synthetic networks that are built to resemble the structure of real men--men sexual contact networks. There are a series of other models, besides the power law, that have been found to fit well the degree distribution of sexual contact networks~\cite{Handcock2004,Hamilton2008}. However, our results are not only robust for power laws with different exponents, but are also qualitatively equivalent for a well mixed population. Accordingly, we would not observe a different phenomenology by considering other generative network models yielding different degree distributions.
%--- EQUILIBRIUM  ---%
\subsection{Conditions for equilibrium}	
	For characterizing the equilibrium of Eqs.~\eqref{eq:Sp}-\eqref{eq:Inp}, we first focus on the condition such that the system can reach the equilibrium state. Can the system reach the equilibrium state if the the transition probabilities are non zero, $\Gamma_{a \rightarrow b} \neq 0$? Intuitively, since the game and disease dynamics are decoupled, it should not be possible. For proofing so, let us consider the following new set of variables:
		\bea
	\label{eq:changeVariables1}
	P^{i} & \equiv& \Ip^i+\Sp^i\\
	NP^i & \equiv& \Inp^i +\Snp^i \\
	\Ip^i & \equiv & \Ip^i \\
	\Inp^i & \equiv & \Inp^i,
	\eea
where the variables $P$ and $NP$ represent the probability of an agent adopting the protection mechanism, respectively not adopting the protection mechanism. With these variables the recurrence relations become:
\begin{widetext}
\bea
	\Delta P^i \hspace{-0.1cm} &=& \hspace{-0.1cm} P_i(t+1)\hspace{-0.1cm}-\hspace{-0.1cm}P_i(t)  = NP^i \Tnpp -P^i \Tpnp  \label{eq:dotP}\\ 
	\Delta NP^i \hspace{-0.1cm} &=& \hspace{-0.1cm} NP^i(t+1)\hspace{-0.1cm}-\hspace{-0.1cm}NP^i(t)  = -NP^i \Tnpp +P^i \Tpnp  \label{eq:dotNP} \\
	\Delta \Ip^i \hspace{-0.1cm} &=&  \hspace{-0.1cm} \Ip^i(t+1)\hspace{-0.1cm}-\hspace{-0.1cm}\Ip^i(t)  = -\mu \Ip^i  +(P^i-\Ip^i) \lambda \gamma (\Ip^i+\Inp^i)-\Tnpp  \Inp^i+\Tpnp  \Ip^i  \\ 
		\hspace{-0.1cm} \Delta \Inp^i \hspace{-0.1cm} &=& \hspace{-0.1cm} \Inp^i(t+1)\hspace{-0.1cm}-\hspace{-0.1cm}\Inp^i(t) \hspace{-0.1cm} = \hspace{-0.1cm} -\mu \Inp^i \hspace{-0.1cm}+\hspace{-0.1cm}(NV^i-\Inp^i)\lambda  (\gamma \Ip^i+\Inp^i)+\Tpnp  \Ip^i -\Tnpp  \Inp^i.
\eea
\end{widetext}
%	\bea
%	\Delta P^i \hspace{-0.3cm} &=& \hspace{-0.3cm} P_i(t+1)\hspace{-0.1cm}-\hspace{-0.1cm}P_i(t)  \\
%	&=& NP^i \Tnpp -P^i \Tpnp  \label{eq:dotP}\\ 
%	\Delta NP^i \hspace{-0.3cm} &=& \hspace{-0.3cm} NP^i(t+1)\hspace{-0.1cm}-\hspace{-0.1cm}NP^i(t)  = -NP^i \Tnpp +P^i \Tpnp  \label{eq:dotNP} \\
%	\Delta \Ip^i \hspace{-0.3cm} &=&  \hspace{-0.3cm} \Ip^i(t+1)\hspace{-0.1cm}-\hspace{-0.1cm}\Ip^i(t)  = -\mu \Ip^i  +(P^i-\Ip^i) \lambda \gamma (\Ip^i+\Inp^i)-\Tnpp  \Inp^i+\Tpnp  \Ip^i  \\ 
%		\hspace{-0.2cm} \Delta \Inp^i \hspace{-0.3cm} &=& \hspace{-0.3cm} \Inp^i(t+1)\hspace{-0.1cm}-\hspace{-0.1cm}\Inp^i(t) \hspace{-0.1cm} = \hspace{-0.1cm} -\mu \Inp^i \hspace{-0.1cm}+\hspace{-0.1cm}(NV^i-\Inp^i)\lambda  (\gamma \Ip^i+\Inp^i)+\Tpnp  \Ip^i -\Tnpp  \Inp^i.
%	\eea
	Since the transition probabilities $\Tnpp $ and $\Tpnp $ contain a Heavyside function $\Theta(\Pp-\Pnp)$ and $\Theta(\Pnp-\Pp)$, respectively (see Eqs.~\eqref{eq:Tpnp} and \eqref{eq:Tnpp}), they cannot be non zero simultaneously. Consequently, the two terms in the first two equations \eqref{eq:dotP} and \eqref{eq:dotNP} cannot compensate each other such that $\Delta P^i = \Delta NP^i = 0$. Therefore, equilibrium can only be reached if $\Tnpp  = \Tpnp  = 0$, i.e. $\Pp = \Pnp$ (see Eq.~\eqref{eq:eqGame}). \\

\end{document}